# A Fast Compressive Channel Estimation with Modified Smoothed L0 Algorithm

Guan Gui[1,2], Wei Peng[2], Qun Wan[1], and Fumiyuki Adachi[2]

1. Dept. of Electric Engineering, University of electrical Science and Technology of China, Chengdu, 611731, China
2. Dept. of Electrical and Communication Engineering, Graduate School of Engineering, Tohoku University, Sendai, 980-8579, Japan

*Abstract*—Broadband wireless channel is a time dispersive and becomes strongly frequency selective. In most cases, the channel is composed of a few dominant coefficients and a large part of coefficients is approximately zero or zero. To exploit the sparsity of multi-path channel (MPC), there are various methods have been proposed. They are, namely, greedy algorithms, iterative algorithms, and convex program. The former two algorithms are easy to be implemented but not stable; on the other hand, the last method is stable but difficult to be implemented as practical channel estimation problems because of computational complexity. In this paper, we proposed a novel channel estimation strategy by using modified smoothed $\ell_0$ (MSL0) algorithm which combines stable and low complexity. Computer simulations confirm the effectiveness of the introduced algorithm comparisons with the existing methods. We also give various simulations to verify the sensing training signal method.

*Keywords- Smooth L0 Algorithm, Phase Transition, Sparse Channel Estimation, Compressive Sensing*

## I. INTRODUCTION

Time dispersive and frequency-selective fading channels often encounter in many communication systems such as mobile wireless channels, indoor radio channels, and underwater acoustic channels [1]. In general, such the multipath propagation channels distort the transmitted signal and channel equalization is necessary at a receiver but many types of equalizers require accurate channel estimation for good performance. In many studies, densely distributed channel impulse response was often assumed. Under this assumption, it is necessary to use a redundant training sequence. In addition, the linear channel estimation methods, such as least square (LS) algorithm, always lead to bandwidth inefficiency. It is an interesting study to develop more bandwidth efficient method to acquire channel state information.

Recently, the compressive sensing (CS) has been developed as a novel technique. It is regarded as an efficient signal acquisition framework for signals characterized as sparse or compressible in time or frequency domain. One of applications of the CS technique is on compressive channel estimation. If the channel impulse response follows sparse distribution, we can apply the CS technique. As a result, the training sequence can be reduced compared with the linear estimation methods. Recent channel measurements show that the sparse or approximate sparse distribution assumption is reasonable [2, 3]. In other words, the wireless channels in real propagation environments are characterized as sparse or sparse clustered; these sparse or clustered channels are frequently termed as a sparse multi-path channel (SMPC). An example of SMPC impulse response channel is shown in Fig.1. Recently, the study on SMPC has drawn a lot of attentions and concerning results can be found in literature [4-6]. Correspondingly, sparse channel estimation technique has also received

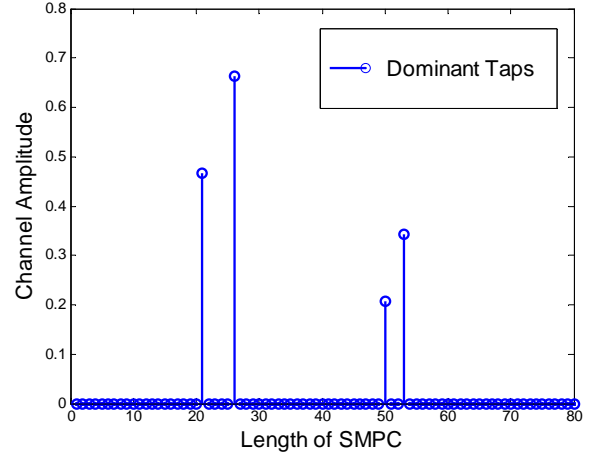

Figure 1. An example of SMPC where the length is 80 while its number of dominant taps is 4.

considerable interest for its advantages in high bit rate transmissions over multipath channel [7].

Exploiting the sparse property of SMPC, orthogonal matching pursuit (OMP) algorithm [8, 9] and convex program algorithm [10] have been proposed. OMP algorithm is fast and easy to be implemented. However, the stability of OMP for sparse signal recovery has not been well understood yet. To mitigate the unstability, Needell and Tropp have presented a compressive sampling matching pursuit (CoSaMP) algorithm for sparse signal recovery in [11]. After that, Gui *et al* [12] have introduced the algorithm to SPMC channel estimation and acquired robust channel estimator. [13]. However, as the number of channel dominant coefficients increasing, accurate channel estimator hard to obtain because of unavoidable correlation between columns of the training sequence, thus the instability of the CoSaMP algorithm easily leads to weak channel estimation.

Convex program method can resolve the instability of CoSaMP algorithm. Convex program algorithm, such as Dantzig Selector (DS) [14], is based on linear programming. The main advantage of convex program method is its stability and high estimation accuracy. The convex problem method can work correctly as long as the RIC conditions are satisfied. However, this method is computationally complex and difficult to be implemented [15] in practical broadband communication systems.

In this paper, we propose an approximate optimal estimator for sparse channel estimation. Because of the idea come from smooth $\ell_0$ (SL0) algorithm [16]. Thus, we term this channel estimation method as MSL0. It has both the advantages of the greedy algorithm and the convex program. In other words, MSL0 algorithm combines low computational complexity and robustness on practical channel estimation.

The rest of the paper is organized as follows. Sparse multipath channel model is presented in Section II. Section III will describe compressive SMPC estimation and the MSL0 channel estimator. In section IV, we will compare the performance of the proposed method with the existing methods by simulations. Finally, conclusions are drawn in Section V.

## II. SPARSE MULTIPATH CHANNEL MODEL

At first, the symbols used in this paper are described as following. The superscript $^H$ stands for Hermite transposition. Bolded capital letters denote a matrix where bolded lowercase letters represent a vector. Notation $|\cdot|$ stands for the absolute value. Norm operator $\|\cdot\|_0$ denotes $L_0$ vector norm, i.e., the number of non-zero entries of the vector; $\|\cdot\|_1$ denotes $L_1$ vector norm, which is the sum of the absolute values of the vector entries. $\|\cdot\|_2$ denotes $L_2$ norm. $\tilde{\mathbf{h}}$ and $\mathbf{h}$ indicate estimate channel vector and actual channel vector, respectively.

We consider single-antenna broadband communication systems, which are often described by frequency-selective baseband channel model. The equivalent baseband transmitted $\mathbf{X}$ and received signals $\mathbf{y}$ is given by

$$\mathbf{y} = \mathbf{Xh} + \mathbf{z} \quad (1)$$

where $\mathbf{X}$ is a complex training signal with Toeplitz structure of $N \times L$ dimensions. $\mathbf{z}$ is the $N \times 1$ complex additive white Gaussian noise (AWGN) with zero mean and variance $\sigma^2$. $\mathbf{h}$ is an $L \times 1$ unknown deterministic channel vector which is given by

$$\mathbf{h}(\tau) = \sum_{i=0}^{L-1} h_i \delta(\tau - \tau_i), \quad (2)$$

where $h_i = h_R + jh_I$ are complex channel coefficients and $\tau$ is a delay spread which sampling length $L$ in baseband channel representation. We define the $\ell_0$ norm of sparse channel vector as

$$\|\mathbf{h}\|_0 = \left\{ \sum_{i=0}^{L-1} |\text{sgn}(h_i)| \right\} = T \quad (3)$$

which denotes the number of dominant taps of the SMPC where $T \ll L$. Where the

$$\text{sgn}|h_i| = \begin{cases} 1 & h_i \neq 0 \\ 0 & h_i = 0 \end{cases} \quad (4)$$

denotes the sign function of $|h_i|$. Suppose that there are $T$ dominant channel taps distributed randomly over the channel. And its complex Gaussian distribution are given by

$$P_\sigma(h) = A\exp\left(-|h|^2 / 2\sigma^2\right). \quad (5)$$

where $|h|$, $h_R$ and $h_I$ represent the module, real and imaginary parts, respectively. And A is a constant which decided by user. From above equation (2-4), we can find the complex channel amplitude $|h| = \sqrt{h_R^2 + h_I^2}$ where its real part $h_R \sim \mathcal{N}(0, \sigma^2)$ and imaginary part $h_I \sim \mathcal{N}(0, \sigma^2)$ are two independent normal distributions.

## III. COMPRESSIVE SMPC ESTIMATION

### A. Optimal estimator

From mathematic perspective, optimal channel estimator can be obtained by compressive sensing (CS) algorithm by

$$\text{minimize } \|\mathbf{y} - \mathbf{Xh}\|_2^2 + \lambda_{optimal} \|\mathbf{h}\|_0. \quad (6)$$

where $\lambda_{optimal}$ is a regularized parameter. Unfortunately, solve the (6) is a non-deterministic polynomial-time (NP) hard problem and thus we should be find suboptimal solution with $\ell_1$ sparse constraint. Such as LASSO estimator is given by

$$\text{minimize } \|\mathbf{y} - \mathbf{Xh}\|_2^2 + \lambda_{LASSO} \|\mathbf{h}\|_1 \quad (7)$$

where $\lambda_{LASSO}$ is a regularized parameter. We can obtain accurate channel estimator from above algorithm (7). However, the compute complexity is very high in (7) and thus hart to implement in practical communication. In the next, we are going to introduce a fast and robust channel estimation method, which term as SL0 algorithm.

### B. MSL0 channel estimator

From above (4), we can built the following function

$$\lim_{\sigma \to 0} C_\sigma(h_i) = 1 - \text{sgn}|h_i|, \quad (8)$$

and therefore by defining $\lim_{\sigma \to 0} J_\sigma(\mathbf{h}) = \sum_{i=1}^{L} C_\sigma(h_i)$, and then we have SMPC measure [15]:

$$\lim_{\sigma \to 0} J_\sigma(\mathbf{h}) \approx \sum_{i=1}^{L} (1 - \text{sgn}|h_i|) = L - \|\mathbf{h}\|_0 \quad (9)$$

And as a result, the dominant coefficient of $\mathbf{h}$ can be approximated by

$$\|\mathbf{h}\|_0 \approx L - J_\sigma(\mathbf{h}). \quad (10)$$

Approximate optimal channel estimator can be given by

$$\text{minimize } \|\mathbf{y} - \mathbf{Xh}\|_2^2 + \lambda_{LASSO}(L - J_\sigma(\mathbf{h})). \quad (11)$$

According to convex optimization theory, we can also change (11) as approximate LASSO and DS problems, respectively. Thus, MSL0 channel estimator given by

$$\text{minimize } (L - J_\sigma(\mathbf{h})) \text{ suject to } \|\mathbf{y} - \mathbf{Xh}\|_2^2 \leq \varepsilon_1 \quad (12)$$

$$\text{minimize } (L - J_\sigma(\mathbf{h})) \text{ suject to } \|\mathbf{X}^T(\mathbf{y} - \mathbf{Xh})\|_\infty \leq \varepsilon_2 \quad (13)$$

Both (12) and (13) are equivalent sparse approximation problem which was proved in [17]. Thus, we can solve (12) or (13) to obtain sparse channel estimator. Where $\varepsilon_1$ and $\varepsilon_2$ are constraint parameters given by users. According to duality of the (12) and (13), they can read as,

$$\text{maximize } J_\sigma(\mathbf{h}) \text{ suject to } \|\mathbf{y} - \mathbf{Xh}\|_2^2 \leq \varepsilon_1 \quad (14)$$

$$\text{maximize } J_\sigma(\mathbf{h}) \text{ suject to } \left\| \mathbf{X}^T (\mathbf{y} - \mathbf{Xh}) \right\|_\infty \leq \varepsilon_2 \quad (15)$$

It is worth noting that the value of $\sigma$ tradeoff accuracy of channel estimator and smoothness of the channel sparse approximation. The smaller $\sigma$ is obtained and the better channel estimator can acquire vice verse. From (6) and (9), minimization of the $\ell_0$ norm is equivalent to maximization of $J_\sigma$ for sufficiently small $\sigma$ in (14) or (15). For small values of $\sigma$, $J_\sigma$ contains a lot of local maxima and it is very difficult to directly maximize this function for very small values of $\sigma$.

### C. Oracle Estimator

To evaluate the MSE performance of channel estimators, it is very meaningful compare their achievements with theoretical performance bound in practical wideband communication systems, then they are approximate optimal and further improvements in these systems are impossible. This motivates the development of lower bounds on the MSE of estimators in the sparse channel estimation. Since the channel vector to be estimated is deterministic, and then we can give a lower bound as for the baseline of MSE. Suppose we know the location set $T = \#\{|h_i| > 0 | i = 0, ..., L-1\}$ of dominant channel taps. Thus, the oracle estimator given by

$$\hat{\mathbf{h}}_{oracle} = \begin{cases} (\mathbf{X}_T^H \mathbf{X}_T)^{-1} \mathbf{X}_T^H \mathbf{y}_T, & T \\ 0, & elsewhere \end{cases}, \quad (16)$$

where $\mathbf{X}_T$ is the partial training signal constructed from columns of training signal $\mathbf{X}$ corresponding to the dominant taps of SMPC vector $\mathbf{h}$. It is noting that we call the oracle estimator as oracle bound in Figure 2~5 in the next part.

## IV. SIMULATION RESULTS AND DISCUSSION

The parameters used in the simulation are listed in Tab. 1. To illustrate the performance of proposed algorithm, Figure.2 shows the MSE of dominant taps by employing LS, Lasso, MSL0 and oracle estimator. The estimation error using mean square error (MSE) evaluation criterion can be defined as:

$$\text{MSE} \triangleq \text{E}\left\{ \left\| \mathbf{h} - \hat{\mathbf{h}}_m \right\|_2^2 \right\}. \quad (17)$$

The computer simulation condition is listed in TABLE 1.

TABLE I. SIMULATION CONDITION

| | *Linear algorithm* | LS |
|---|---|---|
| **Estimation methods** | *Convex optimization* | LASSO |
| | | CoSaMP |
| | | MSL0 |
| **Channel fading** | Frequency-Selective | |
| **Channel length** $L$ | 60 | |
| **Training sequence X** | Complex Toeplitz Structure | |
| **Length of X** | 40 | |

### A. MSE comparison versus channel sparsity

It is obvious that smaller MSE performance means better channel estimator and vice versa. The MSE performance comparisons between the LS, LASSO, MSL0, CoSaMP and oracle bound versus channel sparsity which is shown in Figure 2. When the number of dominant is very small, CoSaMP and LASSO have better MSE performance than MSL0. However, as the number of channel dominant coefficients increasing, MSE performance CoSaMP and LASSO become worse than MSL0. Thus, we can say that MSL0 method more robust than others MSE comparison versus channel sparsity. It is interest to noting that CoSaMP is a iterative algorithm which compute complexity growth rapidly with number of dominant coefficients. Thus, the algorithm for sparse channel estimation has a limitation. In other words, if a channel is very sparse, e.g. the number of dominant coefficients is less than 8 in Figure. 3,

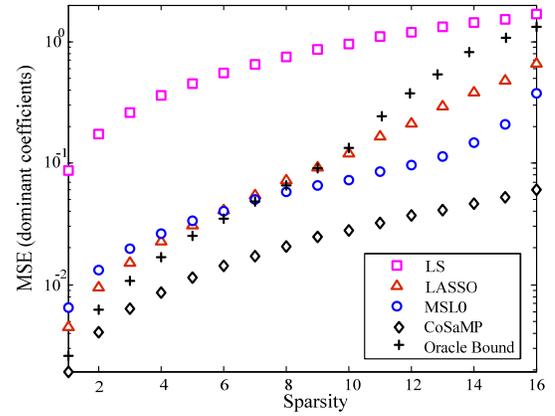

Figure 2. MSE of esimate dominant coefficients with channel sparsity

fast and accurate channel estimator can be acquired and vice versa.

### B. MSE comparison versus SNR

To further study the MSE performance, we compare the different channel estimators versus SRN in Fig. 3. As the SNR increasing, MSL0, LASSO and CoSaMP algorithm approximate achieve the same approximate MSE performance when the number of training sequence keep at constant.

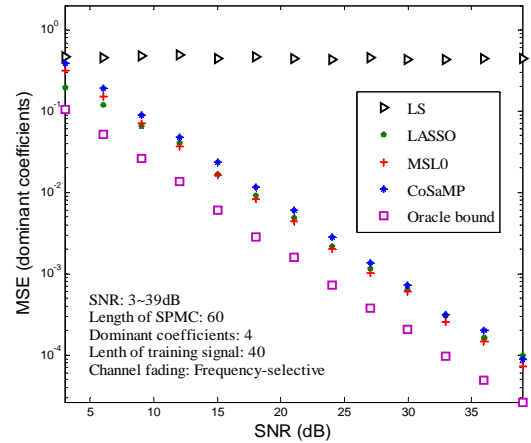

Figure 3. MSE of esimate dominant coefficients with different SNR

*C. Computational Complexity*

To study the computational complexity of the introduced algorithm, we have evaluated the CPU time in second to complete the channel estimation for *SNR*=10*dB*. It is worth mentioning that although the CPU time is not an exact measure of complexity, it can give us a rough estimation of computational complexity. Our simulations are performance in MATLAB 2007 environment using a 2.40GHz Intel Core-2 processor with 2GB of memory and under Microsoft XP 2003 operating system.

The comparison between LS, LASSO, CoSaMP and MSL0 algorithms is shown in Fig. 6. It is seen that the computing time of the algorithm is close to 0.01 seconds for LS and MSL0 algorithm, while the computing time of the LASSO algorithm is more than 0.1 seconds. It is interest to noting that CoSaMP is a iterative algorithm which compute complexity growth rapidly with number of dominant coefficients. Thus, the algorithm for sparse channel estimation has a limitation. In other words, if a channel is very sparse, e.g. the number of dominant coefficients is less than 8 in Figure. 3, fast and accurate channel estimator can be acquired and vice versa.

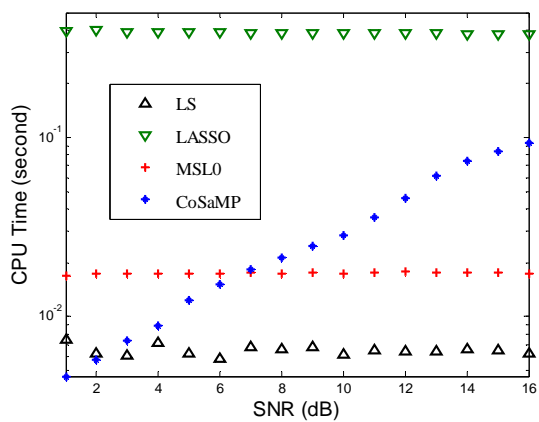

Figure 4. CPU time versus different channel sparsity.

V. CONCLUSION AND FUTURE WORK

In this paper, we have produced a novel sparse channel estimation method MSL0 which combines stable and fast. Thus this method has both advantages of the greedy algorithm and convex program algorithm. It has been shown that, when compared with the existing algorithms, our proposed method is both bandwidth and computationally efficient.

ACKNOWLEDGEMENT

This work is supported in part by the National Natural Science Foundation of China under grant 60772146, the National High Technology Research and Development Program of China under grant 2008AA12Z306, the Key Project of Chinese Ministry of Education under grant 109139 as well as Open Research Foundation of Chongqing Key Laboratory of Signal and Information Processing, Chongqing University of Posts and Telecommunications. And this work is also supported in part by China Scholarship of China Scholarship Council under grant No. 2009607029 and the Outstanding Doctor Candidate Training Fund of University of Electronic Science and Technology of China. And he is also supported in part by Tohoku University Global COE program "Global Education and Research Center for Earth and Planetary Dynamics".

REFERENCE

[1] F. Adachi and E. Kudoh, "New direction of broadband wireless technology," Wirel. Commun. Mob. Comput., vol. DOI: 10.1002/wcm.507, pp. 969-983, May 2007.

[2] Z. Yan, M. Herdin, A. M. Sayeed, and E. Bonek, "Experimental study of MIMO channel statistics and capacity via the virtual channel representation," Tech. Rep., Univ. Winsconsin-Madison, available http://dune.ece.wisc.edu/pdfs/zhou meas.pdf., Feb. 2007.

[3] J. Kivinen, P. Suvikunnas, L. Vuokko, and P. Vainikainen, "Experimental investigations of MIMO propagation channels," Antennas and Propagation Society International Symposium, IEEE, 2002.

[4] W. F. Schreiber, "Advanced television systems for terrestrial broadcasting: Some problems and some proposed solutions," IEEE Proc., vol. 83, pp. 958-981, Jun. 1995.

[5] R. Steele, "Mobile Radio Communications," IEEE Press, 1992.

[6] M. Kocic, D. Brady, and M. Stojanovic, "Sparse equalization for real time digital underwater acoustic communications," roc. OCEANS ⅞ 95, San Diego CA, pp. 1417-1422, Oct. 1995.

[7] C. Carbonelli, S. Vedantam, and U. Mitra, "Sparse channel estimation with zero tap detection," IEEE Transactions on Wireless Communnications, vol. 6(5), pp. 1743-1753, May 2007.

[8] Z. G. Karabulut and A. Yongacoglu, "Sparse channel estimation using orthogonal matching pursuit algorithm," 2004 IEEE 60th Vehicular Technology Conference, vol. 60(6), pp. 3880-3884, 2004.

[9] J. A. Tropp and A. C. Gilbert, "Signal recovery from random measurements via orthogonal matching pursuit," IEEE Transaction on Information Theory, vol. 53(12), pp. 4655-4666, 2007.

[10] U. W. Bajwa, J. Haupt, G. Raz, and R. Nowak, "Compressed channel sensing," To appear in Proc. 42nd Annu. Conf. Information Sciences and Systems (CISS'08), Mar. 19-21, 2008.

[11] D. Needell and J. A. Tropp, "CoSaMP: Iterative signal recovery from incomplete and inaccurate samples," Applied and Computational Harmonic Analysis, vol. 26(3), pp. 301-321., 2008.

[12] [G. Gui, Q. Wan, W. Peng, and F. Adachi, "Sparse Multipath Channel Estimation Using Compressive Sampling Matching Pursuit Algorithm," IEEE APWCS2010, May 19-22, 2010.

[13] E. J. Candès, "The restricted isometry property and its implications for compressed sensing," Compte Rendus de l'Academie des Sciences, Paris, vol. Serie I, 346, pp. 589-592, 2008.

[14] E. Candès and T. Tao, "The Dantzig selector: Statistical estimation when p is much larger than n," Annals of Statistics, vol. 35, pp. 2392-2404, 2006.

[15] N. H. Nguyen and T. D. Tran, "The stability of regularized orthogonal matching pursuit algorithm," http://www.dsp.ece.rice.edu/cs/Stability_of_ROMP.pdf.

[16] H. Mohimani, M. Babaie-Zadeh, and C. Jutten, "Complex-Valued Sparse Representation Based on Smoothed L0 Norm," in proceedings of ICASSP2008, Las Vegas, pp. 3881-3884, April 2008.

[17] M. S. Asif and J. Romberg, "On the LASSO and Dantzig Selector Equivalence," CISS, 2009.